\documentclass[review]{elsarticle}

\usepackage{hyperref,textcomp,amsmath}

\journal{Journal of \LaTeX\ Templates}

\biboptions{sort&compress}

\begin{document}

\begin{frontmatter}

\title{A superfluid liquid helium target for low-momentum electron scattering experiments at the S-DALINAC}

\author{Michaela Hilcker}
\ead{mhilcker@ikp.tu-darmstadt.de}
\author{Jonny Birkhan}
\author{Antonio D'Alessio}
\author{Lars J\"{u}rgensen}
\author{Tobias Klaus}
\author{Peter von Neumann-Cosel}
\author{Norbert Pietralla}
\author{Philipp Christian Ries}
\author{Maxim Singer}
\author{Gerhart Steinhilber}
\address{Institut f\"{u}r Kernphysik, Technische Universit\"{a}t Darmstadt, D-64289 Darmstadt, Germany}

\begin{abstract}
The superconducting electron accelerator S-DALINAC enables electron scattering experiments with low momentum transfer and high energy resolution. In order to perform experiments on helium with high precision and high luminosity, a superfluid liquid helium target with good temperature stability was developed. The functionality of this target could be confirmed and its properties were characterized in a commissioning experiment.
\end{abstract}

\begin{keyword}
helium-4\sep target design\sep S-DALINAC \sep electron scattering
\end{keyword}

\end{frontmatter}


\section{Motivation}
\label{sec:motivation}

\begin{figure}[htb]
	\centering
	\includegraphics[width=0.7\textwidth]{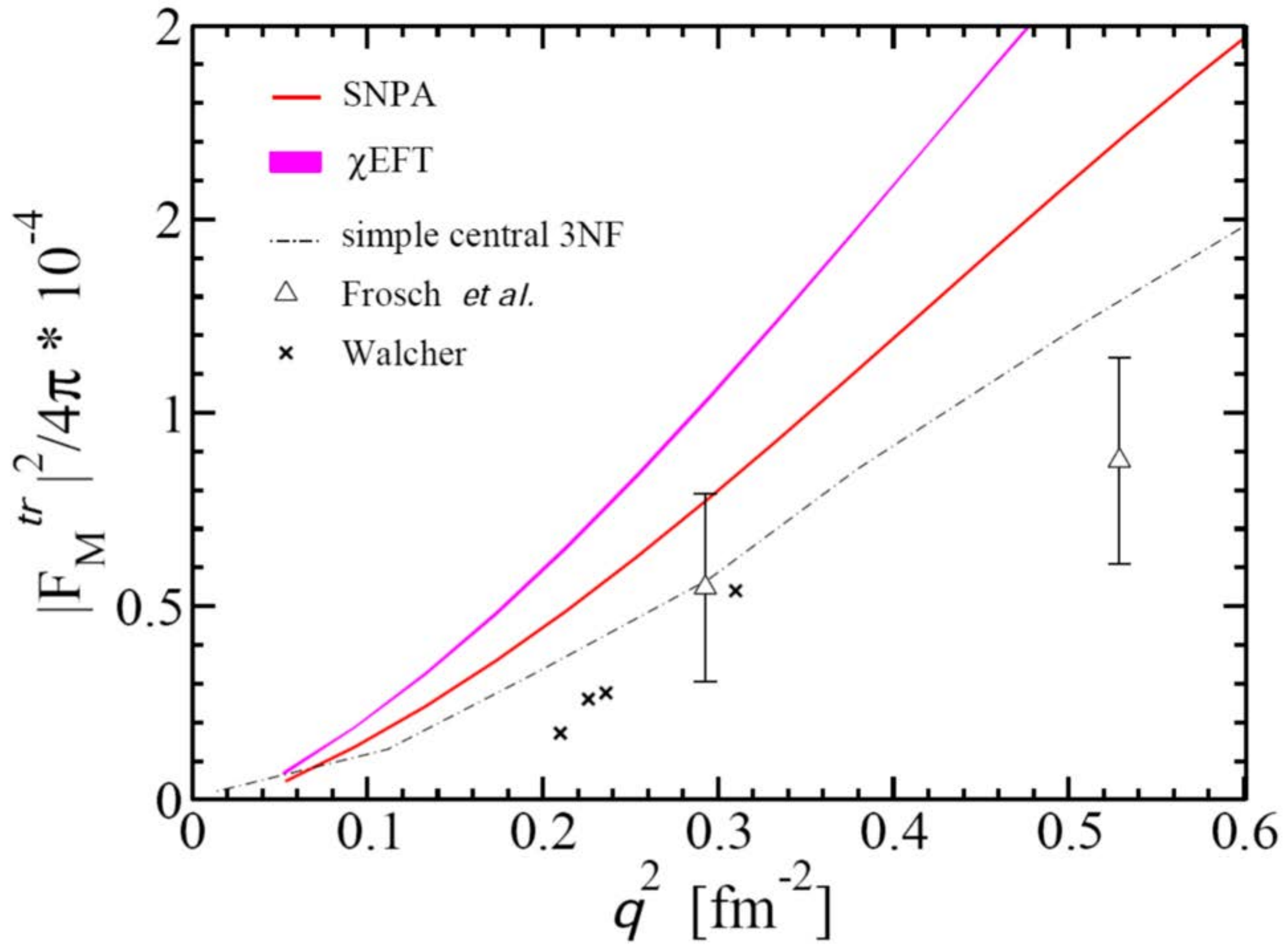}
	\caption{Calculations of $F^{tr}_{M}(q)$ for the ${0}_{1}^{+}\rightarrow {0}_{2}^{+}$ transition in $^{4}$He with the SNPA, the chiral EFT and a simple central 3N force compared to experimental data \cite{Bacca.2013}.}
	\label{fig:bacca1}
\end{figure}

Electromagnetic observables of atomic nuclei such as electron scattering form factors \cite{Hofstadter.1956} are of special interest to test the predictive power of nuclear models because the electromagnetic interaction is very well understood.
In fact, $^{4}$He is an ideal benchmark nucleus to test different dynamical ingredients for the ab initio description of nuclei. In particular, electromagnetic reactions that involve both potentials and external currents can provide new insight. While elastic form factors have been extensively studied and typically the phenomenological standard nuclear physics approach (SNPA) \cite{Carlson.1998} and the more sophisticated chiral effective field theory (chiral EFT) lead to similar results \cite{Bacca.2014}, the situation is quite different for inelastic observables.
A recent ab initio calculation \cite{Bacca.2013} of the monopole transition form factor of $^{4}$He from the ground state to the first excited $0^{+}$ state pointed to a dramatic dependence of the results on the underlying interaction, see Fig.~\ref{fig:bacca1}. Three different Hamiltonians that describe the experimental $^{4}$He ground-state energy within 1\% and lead to the same elastic form factor, show large differences in their predictions for the monopole transition form factor $F^{tr}_{M}(q)$. In particular, at low-momentum transfers, where data were taken in Darmstadt in the 1960s \cite{Walcher.1970}, the discrepancies with respect to the chiral EFT prediction reaches a factor of about three. A new measurement on the first excited $0^{+}$ state of $^{4}$He is needed for concluding on the deviation of the most advanced ab initio calculations from the available data at low momentum transfers.

In order to achieve sufficient luminosities, and thus acceptable measuring times, the previous experiments used normal-fluid liquid targets. A possible problem may be that fluctuating heat input into the target might have led to boiling of the liquid helium and, thus, to bubble formation and an uncontrollable reduction of the effective target thickness. This would have resulted in too small scattering count rates, and hence in underestimated scattering cross sections or form factors.

Therefore we decided to address that issue with a new measurement campaign and, for that, to accept the larger experimental effort of using a superfluid liquid target. The advantage of this approach is that the thermal conductivity of superfluid helium is almost infinitely large, avoiding the production of bubbles and corresponding changes of the effective target thickness.

It is planned to measure the width and form factor of the first excited state of $^{4}$He, the $0^{+}$ state at 20.21 MeV, with a precision of a few per cent at the S-DALINAC by high-resolution inelastic electron scattering at low-momentum transfer using beam energies of 30 - 100 MeV and angular settings ranging from 69\textdegree - 165\textdegree. The data will allow for a precise measurement of the monopole form factor at squared momentum transfers ranging from 1.0 fm$^{-2}$ down to 0.03 fm$^{-2}$. To achieve the desired precision, the scattering signal off the excited state will be measured relative to the elastic scattering cross section off the ground state of $^{4}$He \cite{Sick.2001, Ottermann.1985}, which is known with an accuracy better than 1.5 \%, thereby avoiding many contributions to the systematic uncertainties in an absolute measurement of the cross sections.
Reliable data will be obtained, whose uncertainties will be dominated by counting rate statistics, only.

To achieve sufficient luminosity, a target thickness of about 50 $\text{mg}/\text{cm}^{2}$ is desirable. Since the temperature of the liquid helium has to be reduced in order to accomplish a transition of the fluid into the superfluid phase, the density, and thus the effective target thickness, depends on the exact temperature of the fluid. In order to simultaneously record both the ground state and the excited state at the lowest foreseen beam energy of 30 MeV in a single spectrum, a momentum acceptance of $\pm28 \%$ would be necessary. Since the acceptance of the QClam spectrometer at the S-DALINAC \cite{Pietralla.2018} is limited to $\pm10 \%$, the measurements of the excited state and the ground state must be carried out at two different magnetic field settings. In order to still be able to exploit the advantages of a relative measurement, the temperature stability of the system is crucial for successful experiments.
A large part of the infrastructure required for such measurements is already available at the QClam spectrometer at the S-DALINAC. This provides further geometrical constraints for the final design of the target setup.

Of course, there are several helium targets existing in many different facilities, see e.g. Refs  \cite{Kendellen.2016,Chipps.2014,Kontos.2012,Bass.2009,Ryuto.2005}. But none of them meets our requirements. Therefore we needed to establish a new design.

It is the purpose of this article to describe the design of a superfluid liquid helium (sLHe) target that provides the required constancy of the target thickness and to describe the characterization of the target during a commissioning experiment.
In the next section we will describe the actual realization of the target. Section 3 will present the dedicated cooling procedure which is needed to cool the target down to its working temperature. In section 4, a first measured spectrum and the determination of the effective target thickness will be shown. A summary and a short outlook are given in section 5.

\section{Technical design}
\label{sec:realization}

\begin{figure}[htb]
	\centering
	\includegraphics[width=0.7\textwidth]{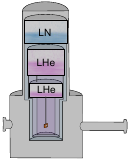}
	\caption{Scheme of the sLHe target. Three cryo tanks, multi-stage heat shields made of aluminum, and the target cell made of copper with aluminum windows are located in an outer vacuum chamber made of stainless steel. There are several holes in the heat shields to allow both the electron beam and the scattered electrons to pass.}
	\label{fig:querschnitt}
\end{figure}

The basic constructive design of the sLHe target is shown in Fig.~\ref{fig:querschnitt}.
It consists of an outer vacuum chamber, three cryo-tanks, the target cell and three actively cooled aluminum heat shields. 
The uppermost tank has a volume of 18.5 liters and is filled with liquid nitrogen. Its main purpose is to cool the outermost heat shield layer. This blocks a large part of the external heat radiation before it can penetrate the liquid helium. In addition, the stability of the system is increased during operation.
The middle tank has a volume of 16 liters and is filled with liquid helium at the boiling temperature of 4.2 K at atmospheric pressure. This tank is used firstly to cool the second heat shield layer to helium temperature and secondly serves as a storage tank for the second helium tank below.
The third tank has a volume of 5.6 liters. It is filled with liquid helium via a needle valve from the upper helium tank. By means of a pump connection, the vapor pressure above the liquid can be reduced by using a feed pump, thereby reducing the temperature of the liquid. At a vapor pressure of 35 mbar, the liquid reaches the lambda point and thus the phase transition to the superfluid phase. During operation, the vapor pressure is reduced to approx. 10 mbar. This corresponds to a temperature of about 1.8 K. At this temperature, the density of superfluid He amounts to $\rho_{\rm He}=0.1453(2)\; {\rm g/cm^{3}}$ and the dependence of the density on the temperature becomes particularly small \cite{Donnelly.1998}. The pressure in the tank is adjusted via a control block connected directly to the pump.

The target cell is supplied with helium from this tank. In addition, the innermost heat shield is cooled such that the temperature stability of the target liquid is further increased.
The target cell is directly connected to the reservoir via a stainless steel tube. Its outer shell is made of copper. Two 0.20(1) mm thick aluminum plates, serving as windows, are clamped between the frames at a distance of $d_{\rm He,0} = 3.7(2)$ mm, sealed with indium. When filled with superfluid helium at a temperature of 1.8 K it provides an expected areal mass density of $\overline{m}_{\rm He,0}=\rho_{\rm He}\cdot d_{\rm He,0}=54(2)\; {\rm mg}/{\rm cm^{2}}$ on perpendicular incidence of the beam. The diameter of the windows is 20 mm. Superfluidity ensures a steady flow of the liquid into the target cell. The unit is fixed with a rotatable CF16 flange at the bottom of the lower helium tank, so that the target cell can be rotated according to the adjusted scattering angle of the spectrometer.

In the heat shields there are a number of circular openings which allow the primary electron beam and the electrons scattered into the spectrometer to pass through the target without additional scattering.
Various measuring instruments continuously record data on the temperature of the individual tanks and their filling levels, respectively. These data are stored in the existing accelerator database by utilizing the EPICS control system \cite{Burandt.2013} and are thus available online during the experiment.

\begin{figure}[h!tb]
	\centering
	\includegraphics[width=0.7\textwidth]{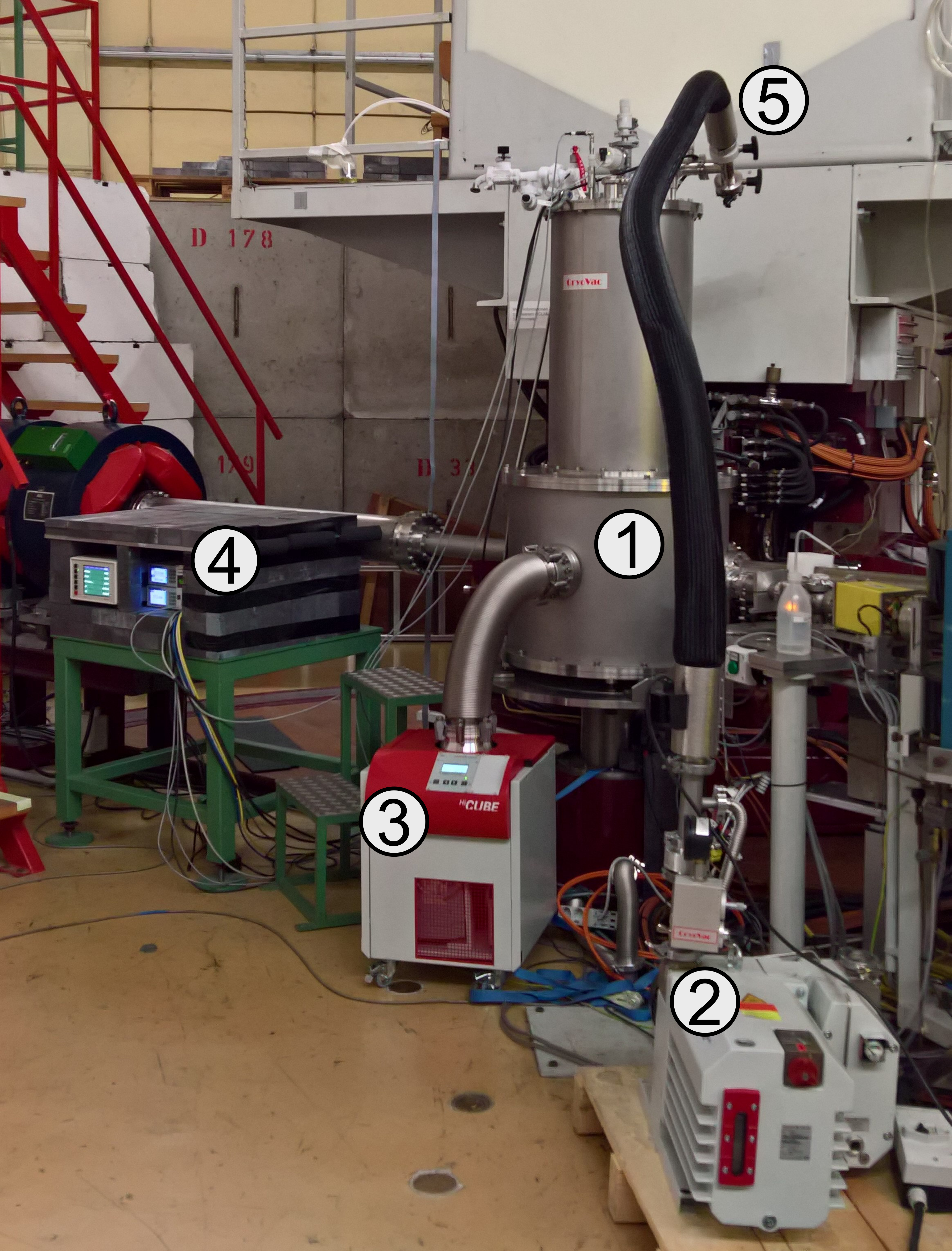}
	\caption{Photography of the assembly for the sLHe target at the QClam spectrometer. The vacuum chamber  with the cryostat and the target cell inside can be seen in the middle \textcircled{1}. The corresponding pump station is placed in front of it \textcircled{3}. A feed pump, at the front right \textcircled{2} is used to cool the liquid helium below the boiling temperature. A self-regulating heating tube warms up the helium before it enters the pump \textcircled{5}. The readout electronics are on the left \textcircled{4}.}
	\label{fig:Foto-QClam}
\end{figure}

A photography of the final assembly is shown in Fig.~\ref{fig:Foto-QClam}. The vacuum chamber with the target and the cryostat inside is in the middle (number 1). In front of it (number 3) is the vacuum pump system. Number 2 shows the feed pump used to cool the liquid helium below the boiling temperature. A self-regulating heating hose (number 5) heats the cold helium gas before it reaches the pump. On the left (number 2) are the readout devices that are connected to the network via Ethernet. The QClam spectrometer is partly visible in the background.

\begin{figure}[h!tb]
	\centering
	\includegraphics[width=0.6\linewidth]{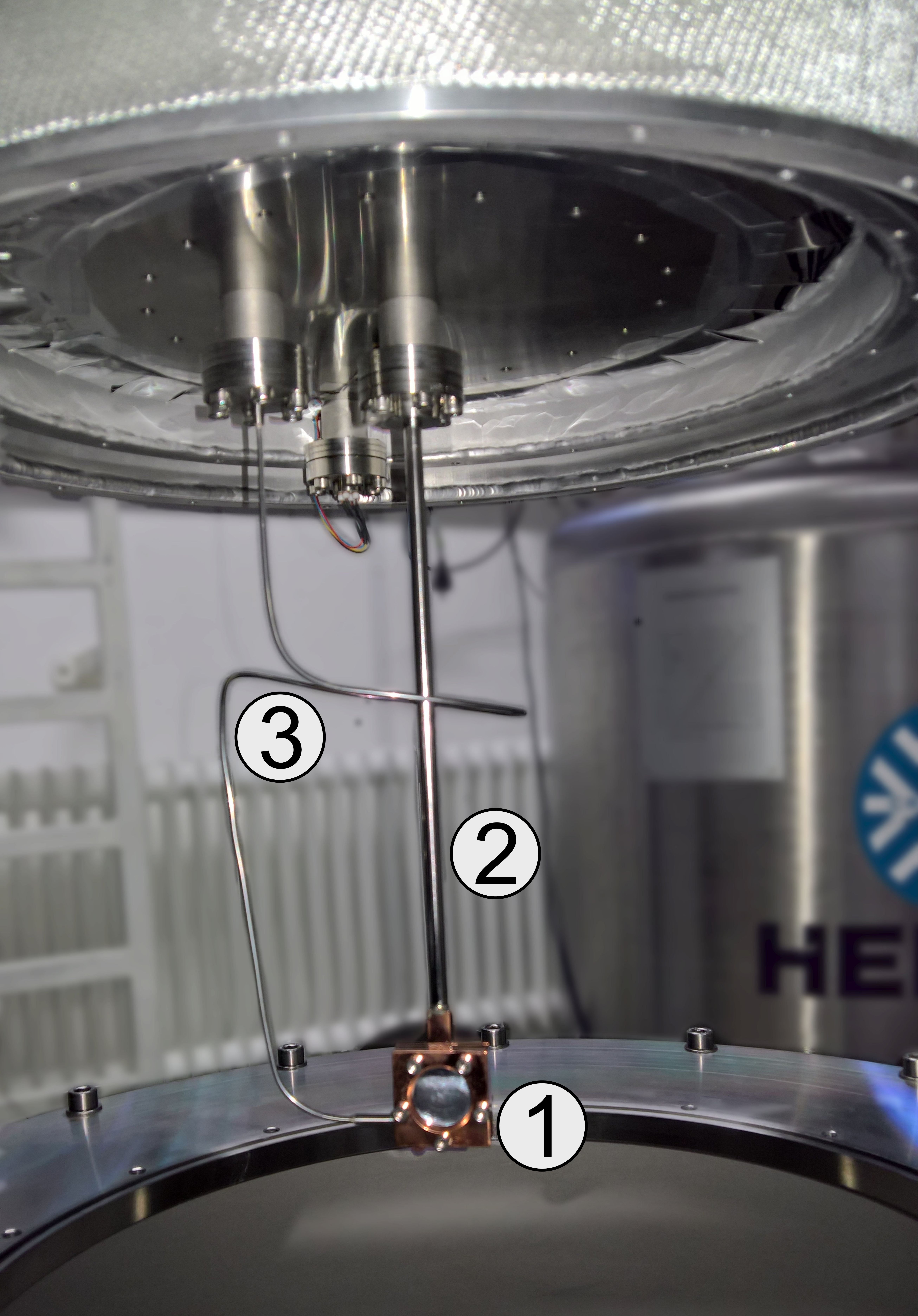}
	\caption{Photography of the target cell. The cell \textcircled{1} consists of a copper shell, which is filled directly from the lower helium tank by a stainless steel tube \textcircled{2}. It can be rotated around the axis of this tube, to match the best imaging conditions of the spectrometer. Two 0.20(1) mm thick aluminum windows are each pressed and screwed between two parts of the copper sheath. Indium is used for sealing. The thinner tube \textcircled{3} ensures unhindered evaporation of the liquid.}
	\label{fig:Foto-Target}
\end{figure}

 A photo of the target cell is shown in Fig.~\ref{fig:Foto-Target}. The tube (number 2) attached to the helium cavity and visible in the middle of Fig.~\ref{fig:Foto-Target}, is directly connected to the bottom of the lower helium tank via a rotatable CF16 flange. The second thinner tube (number 3) is inserted through a second CF flange and ends at the top of the lower helium tank. In this way an unhindered evaporation of the liquid can be ensured. The tube is flexible so that the target can be rotated to the appropriate angle.

\section{Cooling Procedure}
\label{sec:cooling_process}

The target is kept at a constant temperature of 1.8 K during operation. Among other things, the heat shields must be brought to the temperatures specified in Section~\ref{sec:realization}, requiring fast cooling of a large mass. To achieve this, a certain procedure described below has been worked out.

First the whole system is pre-cooled filling the nitrogen tank and then the 4 K helium tank with liquid nitrogen. The needle valve to the lower helium tank must be closed during this step. As soon as there is a sufficient level of liquid in the upper helium tank the needle valve  can be opened carefully allowing liquid nitrogen to drip into the lower tank and evaporate, whereby also the lower tank is cooled. The needle valve must be closed before the boiling temperature of nitrogen is reached, as it will not be possible to remove residual liquid nitrogen from the lower tank.
This procedure must be repeated several times.

Once the entire cryostat reaches stable 90 K, the remaining liquid nitrogen can be pressed out of the upper helium tank.
 When both helium tanks have warmed up again to roughly 100 K, the two helium tanks can then be filled with liquid helium. For this purpose it is important that both tanks are filled at the same time. In addition, the feed pump must be switched on so that there is a slightly lowered pressure in the lower tank.
At the beginning of the filling procedure, the needle valve must be flushed with gaseous helium. To do this, the bypass valve of the control block must be opened and the needle valve needs to be closed at the same time. The pressure in the lower tank drops below 3 mbar. The bypass valve is then closed and the needle valve opened so that as much gaseous helium as possible flows through the needle valve and impurities such as nitrogen or hydrogen are removed. This process is repeated several times.
After that the control block can be adjusted so that a pressure of about 900 mbar is reached in the lower tank.

\begin{figure}[tbh]
	\centering
	\includegraphics[width=0.6\linewidth]{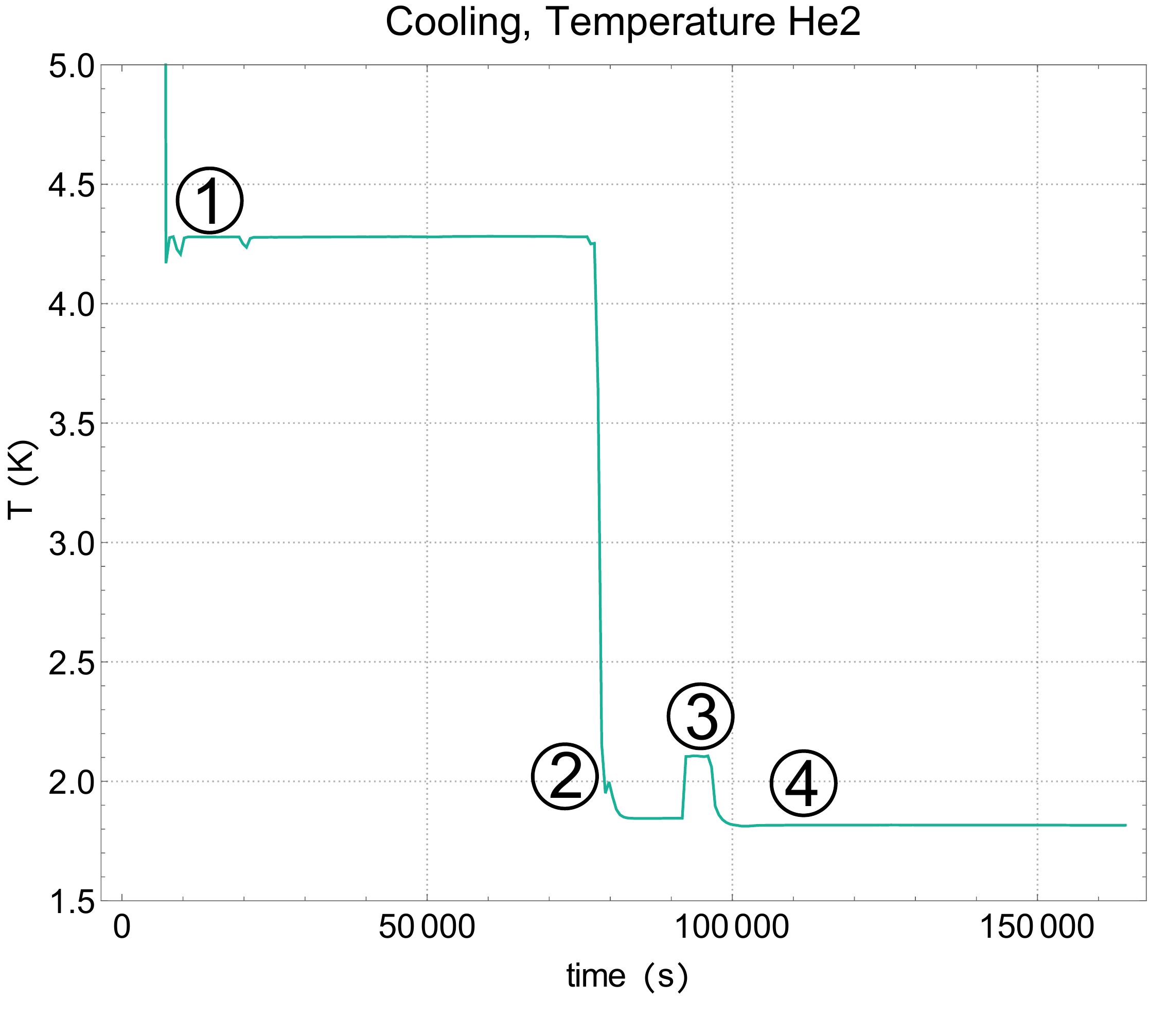}
	\caption{Temperature curve of the lower helium tank during the cooling procedure. \textcircled{1} A liquid level is formed in the tank. \textcircled{2} Phase transition to the superfluid phase. \textcircled{3} The needle valve is opened to increase the time period until the tank needs to be refilled. \textcircled{4} Stable temperature of about 1.8 K reached. The density fluctuation stays below 0.1 \% during experiments.}
	\label{fig:Cooling}
\end{figure}

\begin{figure}[htb]
	\centering
	\includegraphics[width=0.6\linewidth]{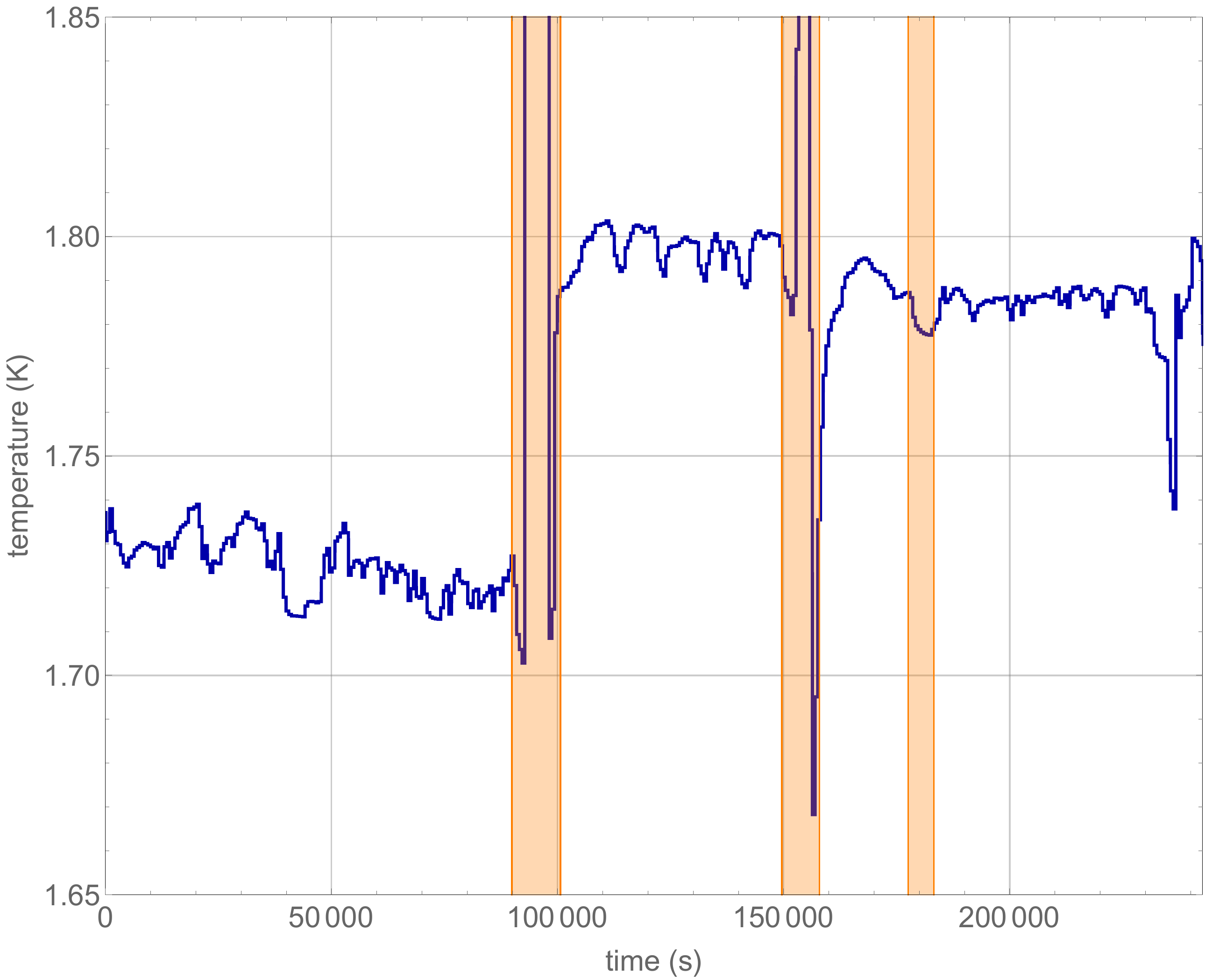}
	\caption{Section of the temperature curve during the experiment. Orange bars represent times during which the system was refilled. During this period no electron beam was present.}
	\label{fig:Temperature}
\end{figure}

As soon as a sufficient level of liquid is reached in both tanks, the temperature in the lower tank, measured using a Cernox CX1030 sensor located at the bottom, can be reduced. The needle valve is closed during this phase of the cooling procedure.
The corresponding temperature curve of the lower helium tank is displayed in Fig.~\ref{fig:Cooling}. 
At number 1 the boiling temperature of helium is reached so that a liquid level can form. For a few hours, the temperature is somewhat unsteady, as the control valve of the feed pump must be adapted to the new situation. Once the temperature is sufficiently stable, the pressure of the vapor above the liquid can be decreased to reduce the temperature of the liquid below the boiling point. At number 2, the lambda point at 2.1 K is finally reached so that the phase transition into the superfluid phase takes place \cite{Donnelly.1998}.
The length of time in which a measurement can be performed without interruption is determined by the evaporation rate of the liquid helium in the lower tank. In order to extend this period, the needle valve is slightly opened again at number 3. Consequently, there is a constant refill from the upper to the lower tank. The control block of the feed pump must be adjusted accordingly. At number 4 a stable temperature of about 1.8 K is finally reached. The pressure above the liquid is roughly 10 mbar.
The temperature of the system was extremely stable during the entire beam time of several days. The temperature fluctuations were less than 0.1 K even when the beam current fluctuated significantly. The density fluctuation thus stays below 0.1 \% \cite{Donnelly.1998} which causes a negligibly small impact on the cross section measurements with an expected precision of a few \%. A section of the temperature curve can be seen in Fig.~\ref{fig:Temperature}.

\section{Target Thickness}

In order to characterize the target properties, a commissioning experiment has been performed at an electron beam energy of 42.5(2) MeV and a scattering angle $\Theta = 75(2)^\circ$, resulting in a momentum transfer of 0.04671(2) fm$^{-2}$. To minimize the contribution of energy loss in the target to the total energy resolution, the target was rotated to align with the bisector between the beam axis and the direction of scattering into the spectrometer, i.e., to an angle $ \theta_{\rm target}=\vartheta/2=37.5(10)^\circ$.

\begin{figure}[htb]
	\centering
	\includegraphics[width=0.8\linewidth]{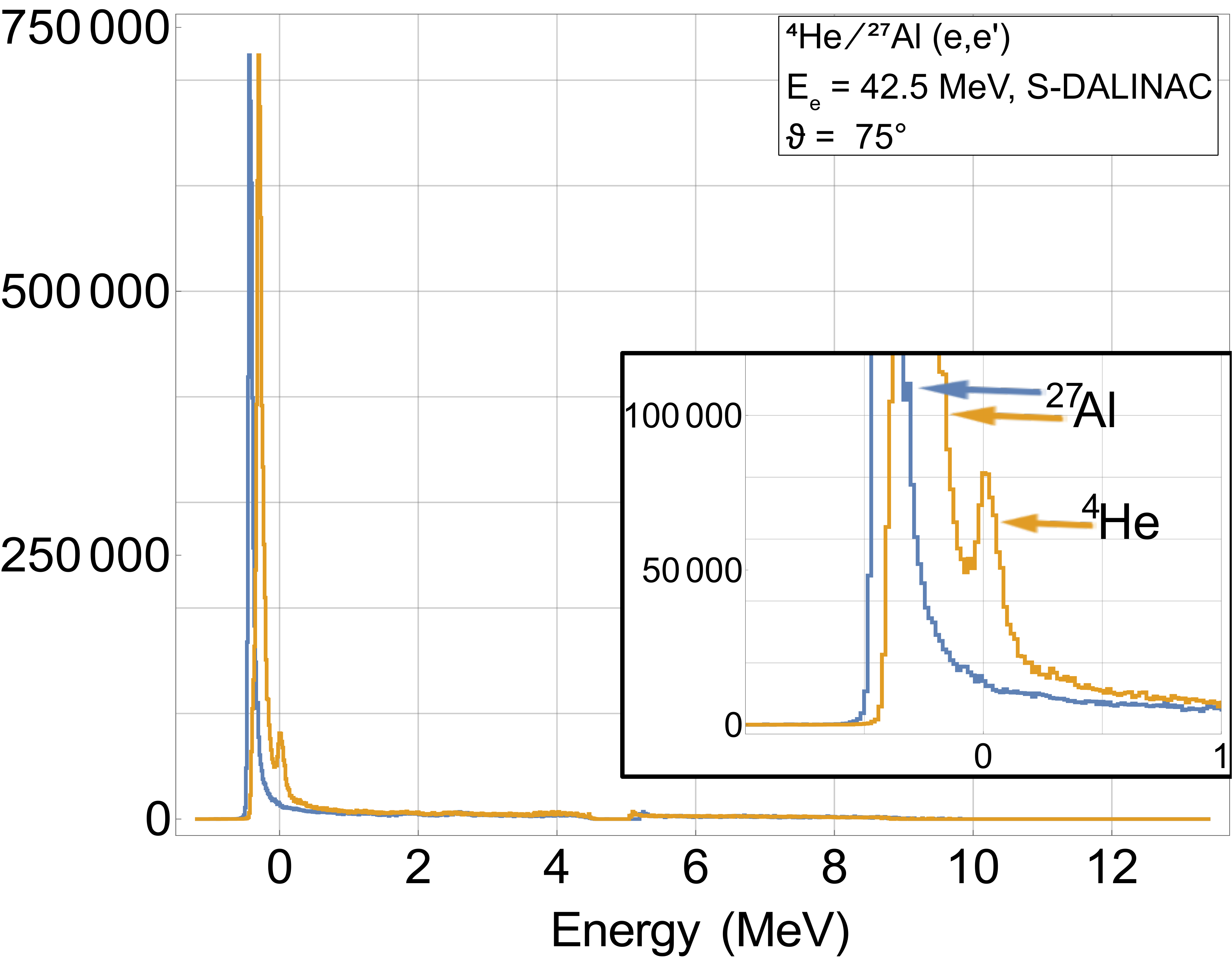}
	\caption{Measured spectrum in the range of elastic scattering. In blue: Background spectrum of the empty target cell. In yellow: Spectrum of the capsule filled with helium. The background spectrum has been normalized to the helium measurement.}
	\label{fig:Spektrum}
\end{figure}

A measured spectrum is shown in Fig.~\ref{fig:Spektrum}, where the inset provides an expanded view around the elastic scattering peak.
The blue line represents a background spectrum. It was obtained by heating the target to 30 K, i.e., above the boiling temperature of helium, in order to determine the contribution from scattering off the aluminum windows.
The yellow line shows the spectrum of the capsule filled with helium. The elastic line from aluminum was normalized to the helium measurement in the figure by comparison of the peak areas of the two lines. This allowed the subtraction of the scattering intensity originating from the aluminum window. The elastic line of aluminum in the helium spectrum is slightly shifted to lower momentum because of the additional energy loss in the liquid helium.
A summary of values specific to the performed experiment can be found in Tab.~\ref{tab:tabelleExp}.

\begin{table}
	\caption{Some specific characteristics of the experiment. The given cross sections have been calculated according to Eq.~(\ref{eq:wq}), see text.}
	\centering
\begin{tabular}{|l|c|r|}
	\hline 
	Beam energy & $E$ & 42.5(2) MeV \\ 
	\hline 
	Scattering angle & $\theta$ & 75(2)° \\ 
	\hline
	Angle of target rotation & $\theta_{\rm target}$ & 37.5(10)° \\
	\hline 
	Peak area helium & $A_{\rm He}$ & 5.02(3) $\times 10^{5}$ \\ 
	\hline 
	Peak area aluminum & $A_{\rm Al}$ & 6.505(55) x 10$^{6}$ \\ 
	\hline 
	Cross section Helium & $\left({\rm d}\sigma / {\rm d}\Omega \right)_{\rm He} $ & 0.049(6) ${\rm mb/sr}$ \\ 
	\hline 
	Cross section Aluminum & $\left({\rm d}\sigma / {\rm d}\Omega \right)_{\rm Al} $ & 2.2(3) ${\rm mb/sr}$ \\ 
	\hline 
\end{tabular} 
	\label{tab:tabelleExp}
\end{table}

We need to consider that the target was rotated by an angle $\theta_{\rm target}$ with respect to the beam axis. Due to this geometric adaption, the electrons had to travel a longer distance within the target than its geometrical thickness $d$.

The integrated areas $A$ of the peaks are proportional to the corresponding scattering cross sections
\begin{align}
&A=Q_{0} \cdot \frac{{\rm d}\sigma}{{\rm d}\Omega}\cdot n = Q_{0} \cdot \frac{{\rm d}\sigma}{{\rm d}\Omega} \cdot \frac{\rho \cdot d_{\rm eff}}{\mu}, \label{eq:peakflaeche}
\end{align}
where $d_{\rm eff} = d/{\rm cos}(\theta_{\rm target})$.
$Q_{0}$ is the integrated collected charge and $n$ the amount of scattering particles per area while $\mu$ denotes the mass of a single atom.
The cross sections
\begin{align}
&\frac{{\rm d}\sigma}{{\rm d}\Omega}=\left ( \frac{{\rm d}\sigma}{{\rm d}\Omega} \right )_{\rm Mott}\cdot \frac{1}{\eta} \cdot\left | F(q) \right |^{2}\label{eq:wq}
\end{align}
for the elastic scattering of aluminum or helium can be calculated using the Mott scattering formula \cite{Hofstadter.1956}.
\begin{align}
&\left ( \frac{{\rm d}\sigma}{{\rm d}\Omega} \right )_{\rm Mott}=\left ( \frac{Z^{2} \alpha^{2}}{4 E^{2} \sin^{4}\left ( \frac{\theta}{2} \right )} \right )\cdot \left ( 1-\beta^{2} \sin^{2}\left ( \frac{\theta}{2} \right ) \right )\label{eq:mott}
\end{align}
Due to the low mass number, the recoil
\begin{align}	
&\eta=1+\frac{2 E}{M \sin^{2}\left ( \frac{\theta}{2} \right )}\label{eq:ruckstoss}
\end{align}
must be taken into account.
The elastic form factors $\left|F(q) \right|^{2} $ in the range of the given momentum transfer for this experiment are sufficiently close to 1 \cite{Sick.2001, Stovall.1967} for the desired precision and thus can be neglected.

We can extract the actual helium thickness $d_{\rm He,exp}$ from our measured spectrum by comparing the ratio of the peak areas of the helium and the aluminum lines with the ratio of the corresponding Mott cross-sections:
\begin{align}
	&\frac{A_{\rm He}}{A_{\rm Al}}=\frac{\left ( \frac{{\rm d}\sigma}{{\rm d}\Omega} \right )_{\rm He}}{\left ( \frac{{\rm d}\sigma}{{\rm d}\Omega} \right )_{\rm Al}}\cdot \frac{\mu_{\rm Al}}{\rho_{\rm Al}\cdot d_{\rm Al}} \cdot \frac{\rho_{\rm He}}{\mu_{\rm He}}\cdot d_{\rm He,exp}. \label{eq:peakverhaeltnis}
\end{align}

Here, the thickness of the aluminum windows is $d_{\rm Al}=(0.400 \pm 0.015) \; {\rm mm}$. $A_{\rm He}$ and $A_{\rm Al}$ are extracted from the spectra.

\begin{figure}[htb]
	\centering
	\includegraphics[width=0.8\linewidth]{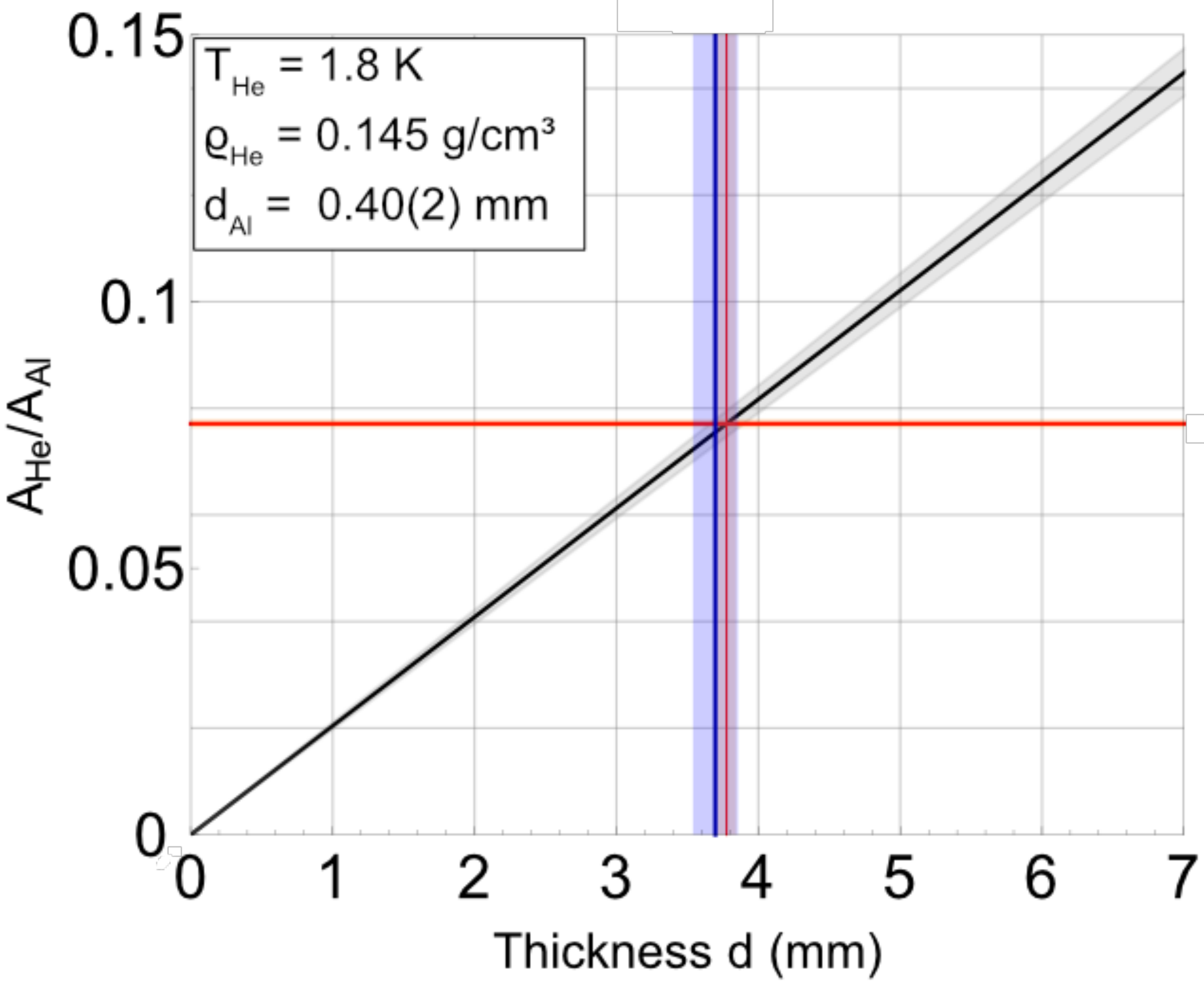}
	\caption{Determination of the experimental target thickness. The black solid line represents the expected ratio of peak areas of elastic electron scattering on helium vs. the aluminum windows as a function of the helium target thickness. The horizontal orange band represents the experimental ratio $A_{\rm He}/A_{\rm Al}$. The intersection point with the black line is used to determine the measured target thickness (vertical orange band). The blue band represents the expected target thickness based upon the manufacturer's specifications.}
	\label{fig:Targetdicke}
\end{figure}

 Using this thickness of the aluminum windows, the black solid line in Fig.~\ref{fig:Targetdicke} represents the calculated ratio of the peak areas as a function of the target thickness $d_{\rm He}$ of the liquid helium. The orange horizontal line results from the measured peak areas. The uncertainty of the peak area ratio is dominated by the uncertainties of the model parameters optimized by the unfolding of the two elastic peaks. The experimental target thickness $d_{\rm He,exp}$ is derived from the intersection of the line representing the ratio of the peak areas with the black line, which comes from Eq.~(\ref{eq:peakverhaeltnis}). The uncertainty is dominated by the manufacturing uncertainty of the aluminum windows. For comparison, the design target thickness $d_{\rm He,des}=3.70(15)$ mm, specified by the manufacturer, is shown in blue. Here the uncertainty is dominated by the unknown thickness of the indium sealing. As can be seen, the experimentally determined thickness conforms with the expected thickness.
As a result, we deduce a target thickness of 3.776 mm with an uncertainty budget as follows: $0.006^{2}\; {\rm mm}^{2}$ variance from counting statistics and $0.044^{2}\; {\rm mm}^{2}$ variance from all other measurands in conformity with the manufacturer's specification. The measurement confirmed an areal mass density of $\overline{m}_{\rm He,exp} = (54.9 \pm 1.0)\; {\rm mg/cm}^{2}$ for sL$^{4}$He, compared to the design value of $(53.8 \pm 2.2)\; {\rm mg/cm}^{2}$.

Another method to determine the experimental target thickness would be the shift of electrons scattered though the Aluminum windows due to the additional liquid He target thickness discussed above. Stopping powers necessary to calculate the energy loss are known for many materials \cite{.21.04.2016} but not for superfluid liquid helium. They may differ from those of LHe due to the possible formation of electron bubbles, in which  shell electrons are pushed aside by additionally entering electrons \cite{Jin.2012}. Therefore, we did not attempt to determine the target thickness through the energy-loss measurement.

\section{Summary}

A superfluid liquid helium target for electron scattering experiments at small momentum transfer was designed and assembled. In addition, a cooling procedure was developed and tested. The obtained temperature stability, is better than 0.1 K even during the entire experiment with a fluctuating electron beam intensity. This leads to a stability of the helium density of better than 0.1 \%. Some key values are summarized in Tab.~\ref{tab:tabelleStatisch} .

The system was successfully tested in a commissioning experiment. To verify the functionality, the target thickness of the superfluid helium was determined experimentally and compared with the manufacturer's specifications. Both values conform with each other.
Thus, the measurement of the first excited state can be started.
It is also planned to use the target in another series of experiments to determine the longitudinal response function for excitation energies above 22 MeV and momentum transfers up to 0.5 fm$^{-2}$ \cite{Bacca.2014}.

\begin{table}
\caption{Key characteristics of the newly established sLHe target at the S-DALINAC.}
\centering
\begin{tabular}{|l|r|}
	\hline
	Helium thickness (Experiment) & 3.78(5) mm \\
	\hline
	Helium thickness (Design value) & 3.70(15) mm \\
	\hline
	Helium mass density (Experiment) & $54.9(10) \frac{\text{mg}}{\text{cm}^2}$ \\
	\hline
	Helium mass density (Design value) & $53.8(22)\frac{\text{mg}}{\text{cm}^2}$ \\
	\hline
	Aluminum windows & $2 \times 0.20(1)$ mm \\
	\hline
	Helium temperature & 1.8(1) K \\
	\hline
	Helium density &  $0.1453(2) \frac{\rm g}{\rm cm^{3}} $\footnotemark{} \\
	\hline
\end{tabular}
\label{tab:tabelleStatisch}
\end{table}
\footnotetext{See Ref.~\cite{Donnelly.1998}}

\section*{Acknowledgments}

We thank the spectrometer group at the IKP for help and all other colleagues at the IKP of TU Darmstadt for experimental shifts. We further thank Michaela Arnold and the entire accelerator group for their support and for providing good electron beams. 
This work was funded by the Deutsche Forschungsgemeinschaft (DFG, German Research Foundation) - Projektnummer 279384907  - SFB 1245.

\bibliography{mybibfile}

\end{document}